\documentclass[11pt,preprint,a4paper]{aastex}
\usepackage{graphics,epsf}
\usepackage{amsmath}                
\usepackage{amsfonts}               
\usepackage{amssymb}                
\usepackage{epsfig}                 

\def \erg{~\rm{erg}}

\begin{document}

\title{\Large{LBV Eruptions Triggered and Powered by Binary Interaction}}

\author{Amit Kashi\altaffilmark{1}}

\altaffiltext{1}{Department of Physics, Technion$-$Israel Institute of
Technology, Haifa 32000 Israel; kashia@physics.technion.ac.il.}


\begin{abstract}
\small{
We suggest that major Luminous Blue Variable (LBV) eruptions are a result of a periastron passage
interaction with the secondary star.
The interaction must take place when the primary envelope is in an unstable phase.
In our model the mass transferred to the secondary accounts for the energy and light curve of the eruption.
We propose that all major LBV eruptions are triggered by stellar companions,
and that in extreme cases a short duration event with a huge mass transfer
rate can lead to a bright transient event on time scales of weeks to months
(a `supernova impostor')}.
\end{abstract}

Luminous Blue Variables (LBVs) are stars in a unique phase of evolution, which occurs only for very massive stars with $M > 50 M_\odot$.
During this short evolutionary phase apart from a continuous mass loss at a large rate,
they undergone a few  major eruptions in which they expel up to few~$\times 10 M_\odot$ of their mass,
and can release energy of up to $\sim 10^{50} \erg$.

The most famous examples for LBV major eruptions are the two 19th century eruption of $\eta$ Car.
This binary system contains a very massive LBV
and a hotter and less luminous main sequence secondary star (Damineli 1996), in a highly eccentric orbit of $e \simeq 0.9$.
The 1837.9 $-$ $\sim$1858 Great Eruption (GE) created
the bipolar Homunculus nebula which contains $10$--$40 \rm{M_\odot}$, and possibly more
(Gomez 2006, 2009; Smith \& Ferland 2007; Smith \& Owocki 2006).
Following the GE, the much less energetic Lesser Eruption (LE) took place between 1887.3-1895.3 (Humphreys et al. 1999),
and only $0.1$--$1 \rm{M_\odot}$ were ejected from the primary (Smith 2005).
A summary of the observed visual magnitude of the eruptions can be found in Frew (2004).

We found that the Lesser Eruption (LE) of $\eta$ Car started close to
the periastron passage of 1887.3 (Kashi \& Soker 2010).
This finding strongly suggests that the interaction with the secondary,
near periastron passage triggered the major eruption.
We generalize this conclusion, and assume that the two luminosity peaks
(rises of more than $1$~mag) of the GE in 1837.9 and 1843
were also triggered by periastron passages.
The mass accreted onto the secondary accounts for the extra energy of the GE.
Not only the energy budget, but also the shape of the light curve, supports the occurrence of mass transfer.
The effect of periastron passage might reveal itself in delaying the decline of the GE and LE, rather than causing a peak in
the light curve (prolongation effect).
The prolongation effect is attributed to the tidal interaction that
perturbed the primary envelope and extended the period of high mass loss rate.
The condition for the prolongation to work is that the primary's envelope
is still in its very unstable phase.

We suggest the following physical process to account for the major LBV Eruptions:
(a) The outer layers of the LBV become unstable due to internal processes unrelated to the secondary.
(b) The periastron passage of the companion exerts tidal forces on the LBV.
(c) The tidal forces amplifies the instability and triggers an eruption, causing the LBV to lose mass.
(d) Part of this mass is accreted by the secondary, and liberates gravitational energy that increases the total luminosity.
In addition, the companion might blow jets that shape the bipolar nebula (Soker 2001).

Solving for the orbital period evolution as a result of mass loss by
the two stars and mass transfer from the primary to the secondary,
which effectively gives a shorter orbital period before the GE,
we could fit the 1837.9 and 1843 sharp rises with the occurrence of periastron passages.
However, in order to achieve that fit we had to use stellar masses much larger
than the previously used values -- the LBV mass is $M_1 = 180$--$200 M_\odot$, and the secondary mass is $M_1 = 70$--$80 M_\odot$.
The mass of the primary before the GE was even larger, $\sim 250 M_\odot$.
These higher than commonly used masses better match the observed
luminosity with stellar evolutionary tracks.
This new finding suggests that $\eta$ Car is one of the most massive binary systems in the Galaxy.
\begin{figure}
\resizebox{0.89\textwidth}{!}{\includegraphics{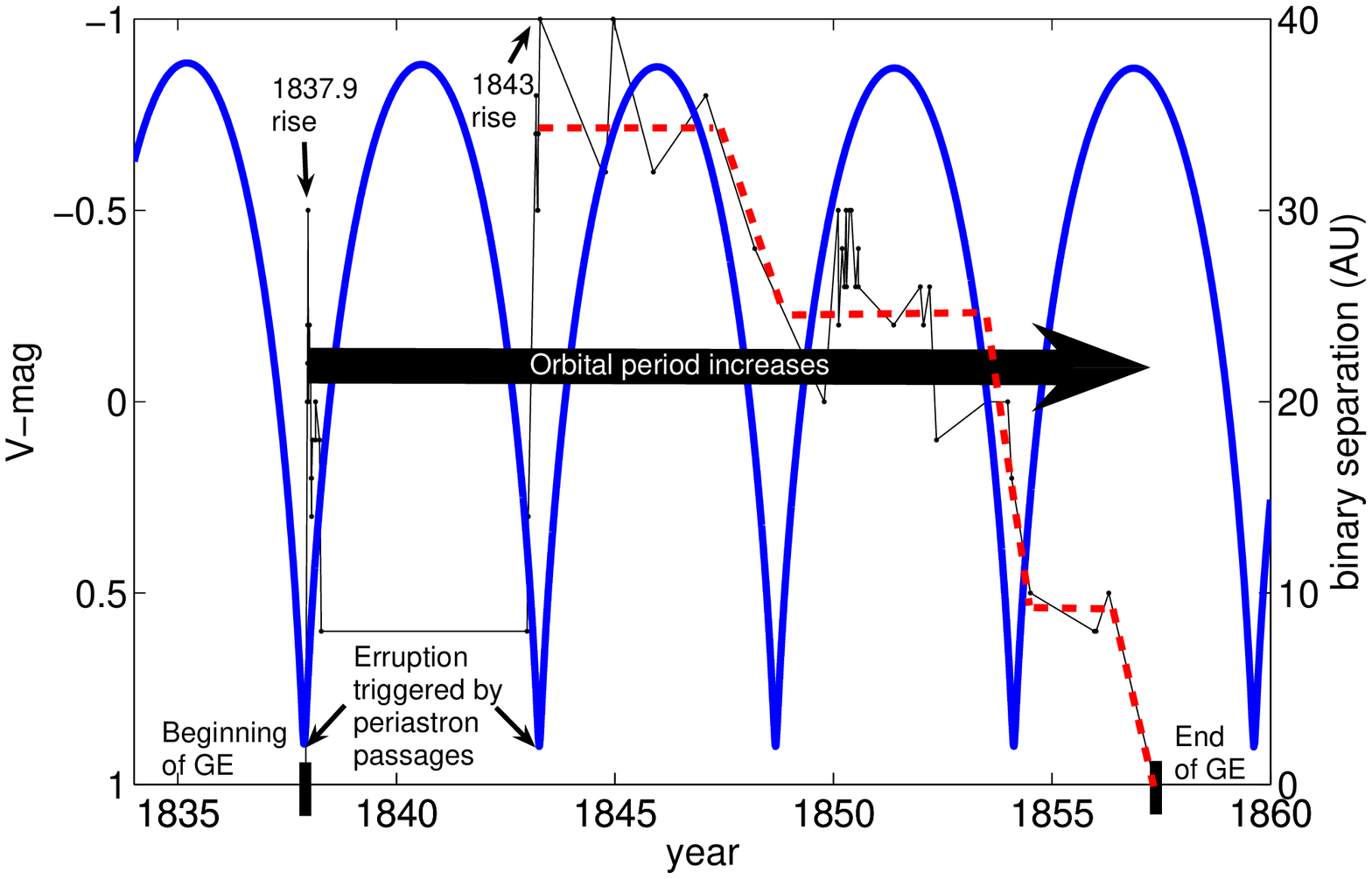}}
\resizebox{0.9\textwidth}{!}{\includegraphics{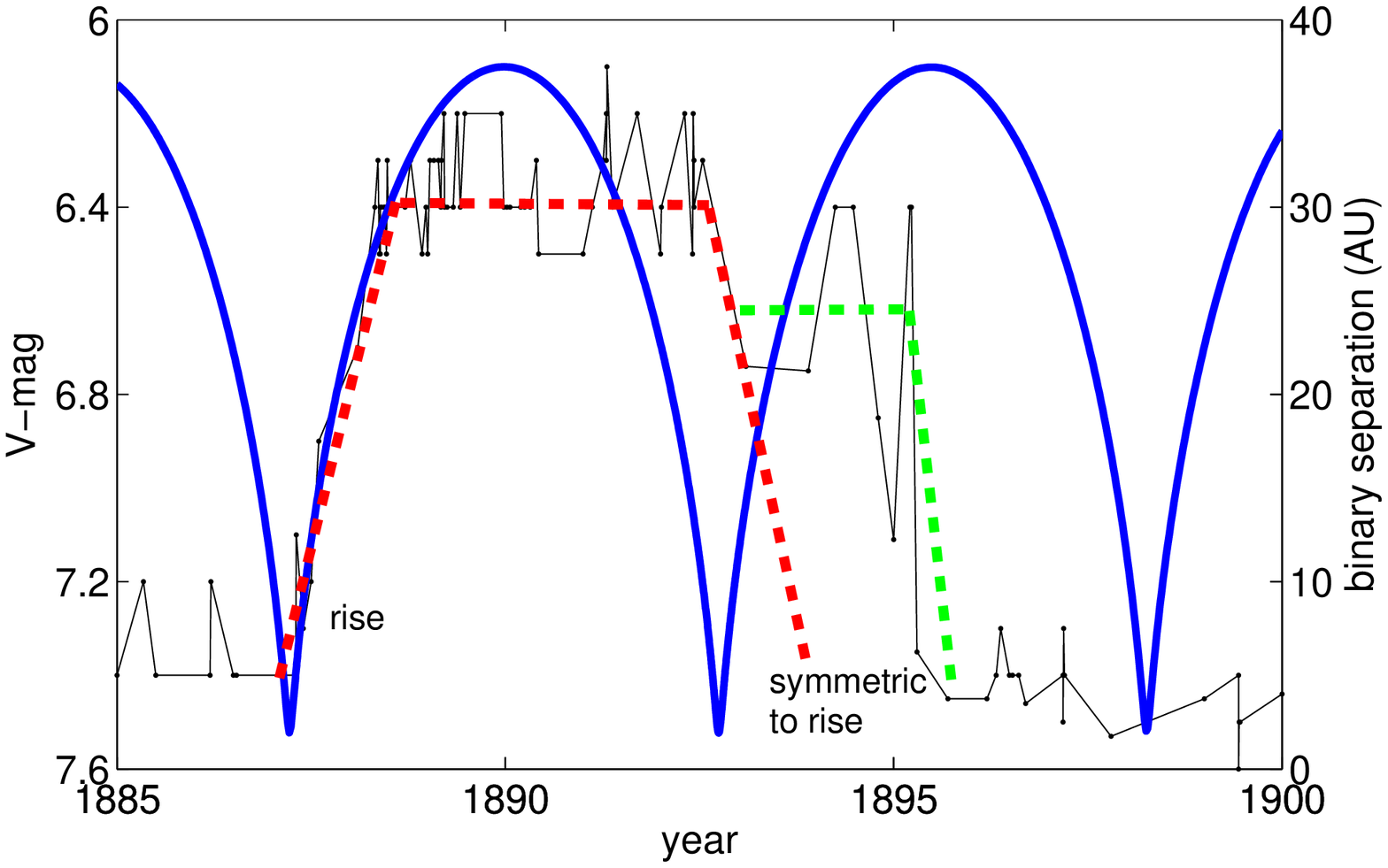}}
\caption{\footnotesize
The binary separation (thick blue line; right axis) and the V-mag light curve (thin black line ;right axis)
during the LE (first panel) and GE (second panel) of $\eta$ Car.
The LE started very close to periastron.
Mass loss and transfer from the primary to the secondary are for the duration of the eruptions.
The masses before the GE are $M_1 = 170 \rm{M_\odot}$ and $M_2=80 \rm{M_\odot}$.
According to our model, during the periastron passage the interaction with the
secondary triggered the eruptions, when the LBV was already in an unstable state.
LE: When we draw a decline line with an opposite slope to that of the rise of the LE,
we find that the light curve behaves as if it was going to end in $1893$--$1894$
(red dashed line), but shortly before then
the periastron passage of 1892.7 took place and prolonged the eruption,
adding another component to the light curve (green dashed line).
GE: The red dashed in line represents the prolongation effect of the last two periastron passages during the GE.
Before each of the last periastron passages the light curve seems as if it starts to rapidly decline.
However, its decline is delayed after the periastron passage, due the tidal interaction near
periastron passage that extends the high mass loss rate phase.
}
\label{fig:MT0}
\end{figure}

Recent simulations of the winds of $\eta$ Car clearly showed that material is accreted onto the secondary
close to periastron passages in present $\eta$ Car (Akashi \& Soker 2010).
All the more so, during the GE accretion must have occurred,
forming an accretion disk around the secondary which blew jets that shaped the Homunculus (Soker 2001).
It is therefore improbable that scenarios in which the orbital plane is not parallel to the symmetry axis of
the Homunculus (e.g., Groh et al 2010a,b) can be correct.

P~Cyg was considered to be a single star, and hence the main possible drawback in a binary interaction model for major LBV eruptions.
To explain the light curve of the 17th century major LBV eruptions of P~Cyg,
showing a series of strong peaks at a decreasing interval (Kashi 2010) (observations from de Groot 1988),
we suggested a similar process -- mass transfer onto a main sequence star.
According to the model the peaks occurred at or very close to periastron passages in a highly eccentric orbit,
when the separation between the stars is considerably smaller than during most of the orbital period.
We find that a mass transfer of $\sim 0.1 M_\odot$ onto a B-type binary companion of $3$--$6 M_\odot$ can account for the
energy of the eruption, and for the decreasing time interval.
In the case of P~Cygni mass transfer was the dominant process (over mass loss) and hence the orbital period was decreasing.
We predicted the companion to have an orbital period of $\sim7$~yrs, and that the Doppler shift should
be possible to detect with high resolution spectroscopic observations.

We propose that all major LBV eruptions are triggered by interaction with stellar companions.
In extreme cases a short duration event with a huge mass transfer
rate can lead to a `supernova impostor', a bright transient event with a very sharp rise in luminosity
and a decay on time scales of weeks to months.

Ironically, with all the modern observational technology no additional detailed enough light curves for major LBV eruptions exist,
and our model is based on observations from hundreds of years ago.
Detailed light curves of LBV eruptions are essential for verification of our model.
Perhaps the series of eruption of the LBV in NGC 3432 (Pastorello 2010) will serve this purpose.

\begin{acknowledgements}
I thank Noam Soker for his major contribution to this research.
\end{acknowledgements}


\begin{references}

\reference{} Akashi, M. \& Soker, N. 2010 (arXiv:1006.3333)

\reference{} Damineli, A. 1996, ApJ, 460, 49

\reference{} de Groot, M. 1988, IrAJ, 18, 163

\reference{} Frew, D. J. 2004, JAD, 10, 6

\reference{} Gomez, H. L., Dunne, L., Eales, S.A., \& Edmunds, M .G. 2006, MNRAS, 372, 1133-1139

\reference{} Gomez, H. L., Vlahakis, C., Stretch, C. M., Dunne, L., Eales, S. A., Beelen, A., Gomez, E. L. \& Edmunds M. G. 2010, MNRAS, 401L, 48

\reference{} Groh, J. H. et al. 2010a, A\&A, 517A, 9

\reference{} Groh, J. H., Madura, T. I., Owocki, S. P., Hillier, D. J. \& Weigelt, G. 2010b, ApJ, 716L, 223

\reference{} Humphreys, R. M., Davidson, K. \& Smith, N. 1999, PASP, 111, 1124

\reference{} Kashi, A. 2010, MNRAS, 405, 1924

\reference{} Kashi, A. \& Soker, N. 2010, ApJ, in press (arXiv:0912.1439)

\reference{} Pastorello, A. et al. 2010 (arXiv:1006.0504)

\reference{} Smith, N. 2005, MNRAS, 357, 1330

\reference{} Smith, N. \& Ferland, G. J. 2007, ApJ, 655, 911

\reference{} Smith, N., \& Owocki, S. P. 2006, ApJ, 645, L45

\reference{} Soker, N. 2001, MNRAS, 325, 584

\end{references}
\end{document}